\documentclass{amsart}
\usepackage{longtable}
\usepackage{amsmath}
\usepackage{graphicx}
\usepackage{xcolor}

\usepackage[numbers]{natbib}
\usepackage[utf8]{inputenc}
\usepackage{array} 
\usepackage{booktabs}
\usepackage{geometry}
\usepackage{mathrsfs}
\usepackage{amsmath}
\usepackage{float}
\usepackage{threeparttable}

\usepackage{amsmath,array}
\usepackage{physics}
\usepackage{multirow}

\def\bfm#1{\mbox{\boldmath$#1$}}
\theoremstyle{definition}

\theoremstyle{remark}

\numberwithin{equation}{section}
\begin{document}

\title[Homogeneity test]{Testing the Equality of Proportions for Combined Unilateral and Bilateral Data}%
\author{Chang-Xing Ma, Kejia Wang}%

\address{Department of Biostatistics, University at Buffalo, New York 14214, USA}%
\email{cxma@buffalo.edu}%
\thanks{Corresponding author: Chang-Xing Ma (cxma@buffalo.edu)}


\begin{abstract}
Measurements are generally collected as unilateral or bilateral data in clinical trials or observational studies. For example, in ophthalmologic studies, statistical tests are often based on one or two eyes of an individual. For bilateral data, recent literatures have shown some testing procedures that take into account the intra-class correlation between two eyes of the same person. \citet{Ma_2013R} investigated three testing procedures under Rosner's model. In this paper, we extend Ma's work for bilateral data to combined bilateral and unilateral data. The proposed procedures are based on the likelihood estimate algorithm derived from the root of 4th order polynomial equations and fisher scoring iterations. Simulation studies are performed to compare the testing procedures under different parameter configurations. The result shows that score test has satisfactory type I error rates and powers. Therefore, we recommend score test for testing the equality of proportions. We illustrate the application of the proposed methods with a double-blind randomized clinical trial.

\end{abstract}
\maketitle
\section{Introduction}
In randomized clinical trials involving paired organs of human body (e.g. eyes, hands, ears), observations are often obtained on one of the paired organ or both of the paired organ of the same individual. For example, in a study of differences between four genetic groups on certain measurements made in a routine ocular examination \cite{Rosner_1982}, a data set was obtained from an outpatient population of 218 persons aged 20-39 with retinitis pigmentosa (RP) who were seen at the Massachusetts Eye and Ear Infirmary. The patients were classified into four genetic types: autosomal dominant RP (DOM), autosomal recessive RP (AR), sec-linked RP (SL)
and isolate RP (ISO). An eye was considered affected if the visual acuity  was 20/50 or worse. 216 patients who had complete information for visual acuity were chosen from the 218 patients. The distribution of the number of affected eyes in the four genetic groups is shown in Table~\ref{intro_exp1}.
\begin{table}
\caption{\label{intro_exp1}Distribution of the number of affected eyes for persons in the four genetic groups}
\begin{tabular}{cccccc}  \hline
&&\multicolumn{4}{c}{Genetic Type}\\\cline{3-6}
number of affected eyes &&  DOM     &  AR  &  SL & ISO   \\\hline
 0 &&  15  & 7 & 3   & 67 \\
 1 &&  6   & 5 & 2   & 24 \\
 2 &&  7   & 9 & 14   & 57 \\\hline
\end{tabular}
\end{table}
If information from both of the paired organs of the same person is available like the data set obtained from the 216 patients in this example, observations from two eyes of a same person are usually correlated \cite{karakosta2011analytic}, thus, standard statistical approaches which assume the independence of observations are not valid and can result in an increased risk of type I error \cite{armstrong2013statistical}. However, as shown in some review articles  \cite{karakosta2011analytic} \cite{glynn2012regression} \cite{murdoch1998people}, statistical methods that take into account the correlation between two eyes are not applied widely and well developed. In this eye example, one may want to investigate whether there is an overall significant difference between the proportion of the affected eyes in the four genetic groups, dealing with correlated data,  \citet{Rosner_1982} proposed models for testing homogeneity of proportions under equal R assumption, where R is a measure of dependence between two eyes of an individual. However, the maximum likelihood estimates and asymptotic testing procedures were not given. \citet{Donner_1989} proposed an alternative approach for testing the equality of proportions based on an adjustment of the chi-square test. Based on these two models, there has been a series of articles on statistical approaches for correlated binary data. \citet{Tang_2008} investigated several procedures for testing the equality of proportions between two groups under Rosner's model. \citet{Ma_2013R}\cite{Ma_2013rho} further derived the maximum likelihood estimate algorithm and investigated several testing procedures for testing equality of proportions under Rosner's model and Donner's model. To compare proportions across groups or strata, three measurements are commonly used: the difference, the relative risk, and the odds ratio \cite{wang2015exact}. \citet{Zhuang_2018}\cite{Zhuang_2018CI} derived several confidence interval (CI) methods for proportion
ratios and several test statistics for testing common ratios of two proportions across strata under the assumption of equal correlation coefficient within each strata. \citet{xue2019interval} proposed CI methods for the ratio of two proportions that are constructed for comparative clinical trials with stratified design under Rosner's model.  \citet{shen2017testing} investigated three homogeneity tests of difference of two proportions for stratified correlated binary data in the basis of equal correlation model
assumption. \citet{tang2016confidence} constructed several CIs for the difference between two correlated proportions in paired-comparison studies with missing observations.

However, in some scenarios participants may refuse to complete the assessment in both eyes or data from one eye may be unavailable for some reasons, resulting in datasets with information from one eye for some individuals and two eyes for others \cite{murdoch1998people}, which can be seen as an extension of the binary correlated data scenarios. An example of combined correlated bilateral data and unilateral data from a clinical trial is shown in Table~\ref{intro_exp2}.
\begin{table*}
\caption{\label{intro_exp2}Distribution of the number of ears without disease at 14 Days}
\begin{threeparttable}
\begin{tabular}{cccc}  \hline
&&\multicolumn{2}{c}{group}\\\cline{3-4}
number of ears being cured  & &  Cefaclor     &  Amoxicillin  \\\hline
 0 &&  14  &  15\\
 1 &&   9   &  3 \\
 2 &&   21   &  13 \\
 total &&   44   &  31 \\\hline 
 0 &&  24   & 39 \\
 1 &&   38   & 27 \\
 total &&   62   & 66 \\\hline
 
 \end{tabular}
 
      \end{threeparttable}
\end{table*}
The clinical trial was conducted to compare cefaclor and amoxicillin for the treatment of acute otitis media with effusion (OME) after the tympanocentesis \cite{Mandel_1982}. 214 children aged 2 months - 16 years underwent unilateral or bilateral tympanocentesis and then were assigned to receive a 14-day course of one of those two antibiotics randomly. The sample used for this analysis consisted of 203 children out of the sample of 214 children with 106 receiving Cefaclor and 97 receiving Amoxicillin. In each of the group, data was obtained from one eye for some persons and two eyes for others (e.g. in Cefaclor group, 44 persons contributed two eyes and 62 persons contributed one eyes). Inference on the effect measures on data set like this requires integrated methods that can be applied to combined unilateral and bilateral samples. So far little work has been done on statistical methods for combined unilateral and bilateral samples. \citet{Pei_2008} investigated several procedures to test the equality of the successful cure rates between two treatments under equal correlation assumption. In their approach, however, maximum likelihood estimates were not derived and the simple estimates were used as an alternative, thus the simulation results may have some deviations.

This article is focused on developing testing procedures to test equality of general $g$ proportions for combined unilateral and bilateral data under Rosner's model, taking into account the between eye correlation. In detail, we consider the observed data as in Table~\ref{intro_exp2} but not limit to two groups. Let $\pi_i$ denote the probability of having response in the $i$th group, for example, the probability of ears being cured in the second example (Table~\ref{intro_exp2}). The equality of $\pi_i$ among different groups is of interest. We generalize likelihood ratio test, wald-type test and score test to handle combined unilateral and correlated bilateral data type. The rest of the article is structured as follows. In Section 2, we derive the maximum likelihood estimates for the parameters under Rosner's model and investigate three methods: Likelihood Ratio test, Wald-type test and Score test. Simulation studies are conducted to evaluate and compare the performance of different tests based on empirical type I error rates and powers in Section 3. Section 4 demonstrates our methodologies by applying an example from a double-blind randomized clinical trial. Some concluding remarks are given in Section 5.
\section{Methods}

In this section, we first introduce the notations and models that will be used throughout the rest sections of this article. Consider comparing $g$ groups of individuals with $m_i$ individuals in the $i$th group that contributing two eyes for the study and $n_i$ individuals in the $i$th group that contributing one eye for the study, $i = 1, \ldots, g$, $M=\sum m_i, N=\sum n_i $ (Table~\ref{tabData}). Let $m_{ti} (t=0, 1, 2)$ be the number of subjects with $t$ responses in the $i$th group who contribute two eyes, $n_{ti} (t=0, 1)$ be the number of subjects with $t$ responses in the $i$th group who contribute one eye, $i = 1, \ldots, g$. Let $S_t (t=0, 1, 2)$ and $N_t (t=0, 1)$ be the number of subjects who have exactly $t$ responses, then

$$S_t = \sum_{i=1}^g  m_{ti},$$ 
$$N_t = \sum_{i=1}^g  n_{ti}.$$

In order to address the between-eye correlation, we use the parametric model proposed by Rosner \cite{Rosner_1982}, which assumes equal dependence between two eyes of the same person across groups:
\begin{equation}
Pr(Z_{ijk}=1)=\pi_i, Pr(Z_{ijk}=1|Z_{ij,3-k}=1)=R\pi_i, \label{conditional_probability}
\end{equation}
where $Z_{ijk}=1$ if the $k$th eye of $j$th individual in the $i$th group has a response at the end of the study, and 0 otherwise, $i=1, 2, \ldots, g$, $j=0,\ldots, m_{i}+n_{i}, k=1,2$. $R$ is a positive constant that measures the dependence between two eyes of the same person. Two eyes from the same individual are completely independent for $R$ = 1 and completely dependent for $R\pi_i$ = 1. From (\ref{conditional_probability}), it is easy to show that $Pr(Z_{ij1}=1, Z_{ij2}=1)=Pr(Z_{ij1}=1|Z_{ij2}=1)Pr(Z_{ij2}=1)=R{\pi_i}^2$, $\mathbb{E}Z_{ij1}=\mathbb{E}Z_{ij2}=\pi_i$, $\mathbb{E}{Z_{ij1}}^2=\mathbb{E}{Z_{ij2}}^2=\pi_i$, $\mathbb{E}Z_{ij1}Z_{ij2}=R{\pi_i}^2$. Then the correlation between two eyes of the same individual for the $i$th group can be calculated as follows:
$$
\rho_i=corr(Z_{ij1}, Z_{ij2}) = \frac{\mathbb{E}Z_{ij1}Z_{ij2}-\mathbb{E}Z_{ij1}\mathbb{E}Z_{ij2}}{\sqrt{\mathbb{E}{Z_{ij1}}^2-({\mathbb{E}Z_{ij1}})^2}\sqrt{\mathbb{E}{Z_{ij2}}^2-({\mathbb{E}Z_{ij2}})^2}}=\frac{R{\pi_i}^2-{\pi_i}^2}{\pi_i-{\pi_i}^2}=\frac{\pi_i}{1-\pi_i}(R-1), i=1, \ldots, g.
$$
Let $\tilde{D}=(m_{01}, m_{11},m_{21},\ldots, m_{0g}, m_{1g}, m_{2g}, n_{01}, n_{11},\ldots,  n_{0g}, n_{1g})$ denote the observed data as shown in Table ~\ref{tabData} . Then for the $i$th group, we have:
$$ (m_{0i},m_{1i},m_{2i})\sim Multinomial(m_i,(R{\pi_i}^2-2\pi_i+1, 2\pi_i(1-R{\pi_i}), R{\pi_i}^2))$$
$$ n_{1i}\sim Binomial(n_i,\pi_i) $$

\begin{table}
\caption{\label{tabData}Frequencies of the number of affected eyes for persons in $g$ groups}
\begin{tabular}{cccccccc}  \hline
&&\multicolumn{4}{c}{group}\\\cline{3-6}
number of affected eyes &&  1     &  2  &  \dots & $g$   && total \\\hline
 0 &&  $m_{01}$   & $m_{02}$ & \dots   &$m_{0g}$ &&$S_{0}$\\
 1 &&  $m_{11}$   & $m_{12}$ & \dots   &$m_{1g}$&&$S_{1}$\\
 2 &&  $m_{21}$   & $m_{22}$& \dots   &$m_{2g}$ &&$S_{2}$\\
 total&& $m_{1}$   & $m_{2}$& \dots   &$m_{g}$ &&$M$ \\ \hline
 0 && $n_{01}$ & $n_{02}$ & \dots   &$n_{0g}$&& $N_{0}$\\
 1 && $n_{11}$ & $n_{12}$& \dots   &$n_{1g}$ && $N_{1}$\\
 total && $n_{1}$ & $n_{2}$& \dots   &$n_{g}$&& $N$\\\hline
\end{tabular}
\end{table}

The log-likelihood can be expressed as follows:
\begin{eqnarray*}
  l(\pi_1, \ldots, \pi_g;  R) &=&  \sum_{i=1}^g [m_{0i}\log \left(R\, {\pi_i}^2 - 2\, \pi_i + 1\right) +  m_{1i}\log\left(2\pi_i(1-R\pi_i)\right) +  m_{2i}\log\left(R\, \pi_i^2\right)]\\
  &+&  \sum_{i=1}^g [n_{0i}\log \left(1- \pi_i\right) +  n_{1i}\log \pi_i] + constant.
\end{eqnarray*}

We now derive the constrained and unconstrained MLEs.

\subsection{ Maximum-likelihood estimates }\hfill\\
The hypotheses to test whether the response rates of the $g$ groups are identical are given as
$$H_0: \pi_1=\pi_2=\cdots =\pi_g\ vs.\  H_1: {\rm some\ of\ the\ } \pi_i {\ \rm are\ unequal}$$

(a) Constrained MLEs\\
Under the null hypothesis, the maximum likelihood estimates of $\pi$ and $R$ can be calculated from
$$
{\partial l\over\partial R}=\frac{S_2}{R} + \frac{{\pi}^2\, S_{0}}{R\, {\pi}^2 - 2\, \pi + 1} + \frac{\pi\, S_{1}}{R\, \pi - 1}=0$$ 
and
$${\partial l\over\partial \pi} = \frac{2\, S_2}{\pi} + \frac{\left(2\, R\, \pi - 2\right)\, S_0}{R\, {\pi}^2 - 2\, \pi + 1} + \frac{\left(4\, R\, \pi - 2\right)\, S_1}{2\, \pi\, \left(R\, \pi - 1\right)}+\frac{N_{1}}{\pi} - \frac{N_{0}}{1-\pi}=0,
$$

The MLEs of $\pi$ and $R$ can be solved by a direct algebra calculation as follows
$$\hat\pi_{H_0}=\frac{A + \sqrt{\mathrm{A^2-6 C}}\, \left(\cos\!\left(\mathrm{\theta}\right) - \sqrt{3}\, \sin\!\left(\mathrm{\theta}\right)\right)}{6\, N}
$$
and

\begin{equation}
\hat R_{H_0}=\frac{2\, N\, \hat\pi_{H_0}^2 + \left( - 2\, M - N_{0} - 3\, N_{1} - S_{1}\right)\, \hat\pi_{H_0} + N_{1} + S_{1}}{\hat\pi_{H_0}\, \left(N_{1} - \hat\pi_{H_0}\, \left(2\, M + N_{0} + 3\, N_{1} - 2\, N\, \hat\pi_{H_0}\right)\right)},
\label{RunderH0}
\end{equation}

where

\begin{eqnarray*}
A&=&N_{0} + 5\, N_{1} + 2\, S_{0} + 3\, S_{1} + 4\, S_{2}\\
C&=&\left(3\, N_{1} + S_{1} + 2\, S_{2}\right)\, S_{0} + N_{1}\left(4\, {N_{1}} + 5\, S_{1} + 6\, S_{2} + 2\, N_{0}\right) +{S_{1}}( {S_{1}} + 3\, S_{2} + N_{0}) + 2\, ({S_{2}} + N_{0}\,) S_{2}\\
\theta&=& \frac13 \arccos\!\left(\frac{18\, A\, C - 2\, A^3   - 108\, N\, N_{1}\, \left(N_{1} + S_{1} + S_{2}\right)}{2\, \sqrt{ {\left( A^2-6\, C \right)}^3}}\right)
\end{eqnarray*}

Here, $\hat\pi_{H_0}$ and $\hat R_{H_0}$ are the MLEs under null hypothesis $H_0: \pi_1=\pi_2=\cdots =\pi_g$ for the situation of combined unilateral and correlated bilateral data. The results can also be used for other data type, for example, when $N_0=N_1=0$, the situation is reduced to bilateral data scenario, the MLEs of $\pi_i's$ and $R$ can be simplified as $\frac{S_1+2S_2}{2M}$ and $\frac{4MS_2}{(S_1+2S_2)^2}$, respectively. For unilateral data, i.e., $S_0=S_1=S_2=0$, the data are collected only from one eye of each person, so there is no need to consider the estimate of $R$, the MLEs of $\pi_i's$ can be simplified as $\hat\pi_{H_0}=\frac{N_1}{N_1+N_0}$.\\

(b) Unconstrained MLEs\\
Differentiating $ l(\pi_1, \ldots, \pi_g;  R)$  with respect to parameters $\pi_i$'s and $R$ we have

\begin{equation}
\frac{\partial l}{\partial \pi_i} = \frac{2\, m_{2i}}{\pi_{i}} + \frac{\left(2\, R\, \pi_{i} - 2\right)\, m_{0i}}{R\, {\pi_{i}}^2 - 2\, \pi_{i} + 1} + \frac{\left(4\, R\, \pi_{i} - 2\right)\, m_{1i}}{2\, \pi_{i}\, \left(R\, \pi_{i} - 1\right)} + \frac{n_{1i}}{\pi_i} - \frac{n_{0i}}{1-\pi_i},
\label{eq:deril1}
\end{equation}

\begin{equation}
\frac{\partial{l}}{\partial R}(\pi_1, \ldots, \pi_g; R) =
\frac{S_2}{R} + \sum_{i=1}^g \left[ \frac{{\pi_{i}}^2\, m_{0i}}{R\, {\pi_{i}}^2 - 2\, \pi_{i} + 1} + \frac{\pi_i\, m_{1i}}{R\, \pi_i - 1}\right]
\label{eq:deriR}\end{equation}

The maximum likelihood estimates of $\pi_i$'s and $R$ are the solutions of the equations

%Denote $\hat \pi_i, i=1, \ldots, g$ and $\hat R$ as the maximum likelihood estimates of $\pi$'s and $R$, respectively. $\hat \pi_i$'s and $\hat R$ are the solution of the following equations

\begin{equation}
{\partial l\over\partial \pi_1} = 0,\ \ \ \ \  \cdots \ \ \ \ \   {\partial l\over\partial \pi_g} = 0, \ \ \ \ \ {\partial l\over\partial R}=0.
\label{eq:deril}\end{equation}

There is no close form solution of $(\pi_1, \ldots, \pi_g; R)$ in Equation (\ref{eq:deril}), so it has to be solved iteratively. 

${\partial l\over\partial \pi_i} = 0$, $i=1, \ldots, g$ can be simplified as the following 4th order polynomial

\begin{equation}
a \pi_i^4 + b \pi_i^3 + c \pi_i^2 + d \pi_i + e=0,
\label{4thpoly}\end{equation}

where
\begin{eqnarray*}
a &=& R^2(2m_i+n_i)\\
b &=& -R((4,5,6,3,3)+R(2,2,2,0,1)) D_i\\
c &=& R(4,7,8,1,4)D_i+2N+2m_{2i}\\
d&=&-(2, 3+2R, 6+2R, 1, 3+R) D_i\\
e &=& m_{1i}+2 m_{2i}+ n_{1i}\\
D_i&=& ( m_{0i},  m_{1i},  m_{2i},  n_{0i},  n_{1i})^T.
\end{eqnarray*}
Here, we apply the Fisher scoring method to update $R$ with a given $\pi_i$, which can be obtained from the real root of the above 4th order polynomial. The iteration procedure is described as follows:

(1) Set the initial value of $R$ as $\hat R^{(0)}=\hat R_{H_0}$, where $\hat R_{H_0}$ is the MLE of $R$ under null hypothesis as shown in Equation (\ref{RunderH0}).

(2) Under $\hat R^{(t)}$, obtain $\hat \pi_1^{(t)},\ldots,\hat \pi_g^{(t)}$ from the real root of Equation (\ref{4thpoly}) directly.

(3) The $(t+1)$th update for $R$ can be updated by Fisher scoring method
$$
\hat R^{(t+1)}=\hat R^{(t)}- \left({\partial^2 l\over\partial R^2}(\hat \pi_1^{(t)},\ldots,\hat \pi_g^{(t)}; \hat R^{(t)})\right)^{-1}{\partial l\over\partial R}(\hat \pi_1^{(t)},\ldots,\hat \pi_g^{(t)}; \hat R^{(t)}).
$$
where
$$
\frac{\partial^2{l}}{\partial R^2}(\pi_1, \ldots, \pi_g; R) =
-\frac{S_2}{R^2} - \sum_{i=1}^g \left[ \frac{{\pi_{i}}^4\, m_{0i}}{(R\, {\pi_{i}}^2 - 2\, \pi_{i} + 1)^2} + \frac{\pi_i^2\, m_{1i}}{(R\, \pi_i - 1)^2}\right]
$$
See  Equation~\eqref{eq:deriR} for ${\partial l\over\partial R}(\hat \pi_1^{(t)},\ldots,\hat \pi_g^{(t)}; \hat R^{(t)})$.
%With the MLEs derived,\textcolor{red}{ we consider the following test statistics.}

(4) Repeat step 2 and step 3 until convergence (when $|\hat R^{(t+1)}-\hat R^{(t)}|$ is sufficiently small, say less than $10^{-5}$ ), stop and return the estimates. Denote $\hat \pi_i, i=1, \ldots, g$ and $\hat R$ as the maximum likelihood estimates of $\pi$'s and $R$ under alternative hypothesis, respectively. 

Based on the constrained and unconstrained MLEs, now we derive three test statistics for combined unilateral and correlated bilateral data in the following subsections.

\subsection{Likelihood ratio test ($T^2_{LR}$)}\hfill\\
The likelihood ratio (LR)  test is given by
$$T^2_{LR}=2[l(\hat\pi_1,\ldots,\hat\pi_g; \hat R)-l(\hat\pi_{H_0},\ldots,\hat\pi_{H_0}; \hat R_{H_0})].$$
Under the null hypothesis, $T^2_{LR}$ is asymptotically distributed as a chi-square distribution with $g-1$ degrees of freedom.

\subsection{Wald-type test ($T_W^2$)}\hfill\\
Let ${\bfm \beta}=(\pi_1, \cdots, \pi_g, R)$ and
$$ 
C=\begin{bmatrix}
    1 & -1 &  &  & & 0 \\
    & 1 & -1 &  & & 0 \\
     &  & \ddots & \ddots & & \vdots \\
     & & & 1 &-1 & 0 \\
  \end{bmatrix},
$$
the null hypothesis $H_0: \pi_1=\cdots=\pi_g$ can be alternatively expressed as $H_0: C {\bfm \beta}^T=0$. Then, the Wald-type test statistic ($T^2_W$) for testing $H_0: C {\bfm \beta}^T=0$ can be written as
$$
T^2_W=({\bfm \beta}C^T) (C I^{-1} C^T)^{-1} (C{\bfm \beta}^T) | {\bfm \beta} = (\hat\pi_1, \ldots, \hat\pi_g, \hat R),
$$
where $I$ is the Fisher information matrix for {\bfm \beta} (See Appendix for the formula of the inverse of the information matrix $I^{-1}(\pi, R)$). For simiplity, let 

\begin{eqnarray*}
  D_{ij} &=&  \begin{cases}
 (h a_i-b_i^2)\sum_{k\ne i}a_k-a_i(\sum_{k\ne i}b_i)^2, & \text{if $i=j$},\\
    \sum b_k(b_ia_j+b_ja_i) - ha_ia_j -b_ib_j\sum a_k, & \text{if $i\ne j$},
  \end{cases}\\
    a_i &=& \frac{2\, m_i\, \left(2\, \hat R^2\, {\hat\pi_i}^2 - \hat R\, {\hat\pi_i}^2 - 2\, \hat R\, \hat\pi_i + 1\right)}{\hat\pi_i\, {\left(\hat R\, {\hat\pi_i}^2 - 2\, \hat\pi_i + 1\right)}{\left(1 - \hat R\, \hat\pi_i\right)}}
 + \frac{n_{i}}{\hat\pi_i(1-\hat\pi_i)}, \\
b_i&=& -\frac{2\, \left(1 - \hat R\right)\, {\hat\pi_i}^2\, m_i}{{\left(\hat R\, {\hat\pi_i}^2 - 2\, \hat\pi_i + 1\right)}{\left(1 - \hat R\, \hat\pi_i\right)}}, \\
 h&=&  \sum_{i=1}^g\frac{{\hat\pi_i}^2\, m_i (\hat R\pi_i - 2\, \hat\pi_i + 1)}{\hat R{\left(\hat R\, {\hat\pi_i}^2 - 2\, \hat\pi_i + 1\right)}{\left(1 - \hat R\, \hat\pi_i\right)}}.
\end{eqnarray*}

%$$T_W=\frac{\sum_{i,j=1}^g \hat\pi_i \hat\pi_j D_{ij}}{\sum_{k=1}^g(b_k^2-h a_k)},$$

Then, we have $$T^2_W=\frac{\sum_{i,j=1}^g \hat\pi_i \hat\pi_j D_{ij}}{\sum_{k=1}^g(b_k^2-h a_k)},$$ which is asymptotically distributed as a chi-square distribution with $g-1$ degrees of freedom. Similarly, by choosing other $C$ matrix in the above statistic, we can have other multivariate tests of $\pi_i$'s.
For example, let $c=(0,\ldots,1,\ldots,-1,\ldots,0)$ with 1 in $i$th element and $-1$ in $j$th element, Wald-type test statistic for testing $H_{0a}: \pi_i=\pi_j$ vs $H_{1a}: \pi_i \ne \pi_j, i\ne j$ can be given as
$$
T^2_{Wa}=({\bfm \beta}c^T) (c I^{-1} c^T)^{-1} (c{\bfm \beta}^T) | {\bfm \beta} = (\hat\pi_1, \ldots, \hat\pi_g, \hat R),
$$
which can be simplified as
$$
T^2_{Wa}(i,j)=\frac{a_ia_j(\hat\pi_i-\hat\pi_j)^2(\sum_{k=1}^g(h-b_k^2/a_k))}{(a_i+a_j)(\sum_{k\ne i,j}^g(b_k^2/a_k)-h)+(b_i+b_j)^2},
$$
$T^2_{Wa}$ is asymptotically distributed 
as a chi-square distribution with $1$ degree of freedom.

\subsection{Score test ($T^2_{SC}$)}\hfill\\
Let $$
U\doteq(U_1,\dots,U_g,0)=\left({\partial l\over\partial \pi_1}, \ldots, {\partial l\over\partial \pi_g}, 0\right),
$$
the score test statistic $T_{SC}$ can be expressed as
$$T_{SC}^2=U I(\pi, R)^{-1} U^T | \pi_1=\cdots=\pi_g=\hat \pi_{H_0}, R=\hat R_{H_0}$$
where $I$ is the Fisher information matrix for {\bfm \beta} (See Appendix for the formula of the inverse of the information matrix $I(\pi, R)^{-1}$).
After lengthy algebra calculations, $T_{SC}^2$ can be simplified as
\begin{equation}\label{scoretestS}
T_{SC}^2=\sum_{i=1}^g\frac{U^2}{I_{ii}} + \left(\sum_{i=1}^g\frac{I_{i,g+1}U_i}{I_{ii}}\right)^2
 \left(I_{g+1,g+1}-\sum_{k=1}^g \frac{I_{k,g+1}^2}{I_{kk}} \right)^{-1}
\end{equation}
See Appendix for the formula of $I_{ii}$ and $I_{i,g+1}$.

\section{Monte Carlo simulation studies}

In this section, we perform Monte Carlo simulations to investigate the performance of the proposed testing procedures discussed in Section 2. We also provide a comparison to Donner's adjusted Chi-square approach, which is defined in \cite{Donner_1989}. For Donner's adjusted Chi-square approach, the adjusted Pearson Chi-square statistics with $g-1$ degrees of freedom is given by
$$X^2=\sum_{i=1}^g \left(\frac{(A_i-M_i\hat{\theta})^2}{M_i\hat{\theta}}+\frac{(M_i-A_i-M_i\hat{Q})^2}{M_i\hat{Q}}\right)^2$$
where $A_i=m_{1i}+n_{1i}+2m_{2i}$, $M_i=2m_i+n_i$, $\hat{\theta}=\sum A_i/\sum M_i$, $\hat{Q}=1-\hat{\theta}$. The robustness of the proposed testing procedures are evaluated by empirical type I error rate and power. In section 3.1, we describe the data generating process and the design of the simulations. In section 3.2, we summarize and interpret the findings from the simulations in section 3.1.   
\subsection{Design of the simulations}\hfill\\
In this section, we examine the empirical performance of the proposed methods. A detailed description of the simulation process is in the following subsections. Section 3.1.1 and Section 3.1.2 describe the simulation process for computing empirical type I error. In Section 3.1.1, we consider some specific parameter settings. In Section 3.1.2, we extent such settings to parameters from the whole parameter space. In Section 3.1.3, we show simulation process for computing the power under some specific parameter settings.
\subsubsection{Empirical type I error under specific parameter settings}\hfill\\ 
%\subsubsection{Empirical type I error under specific parameter settings}\hfill\\
First, we conduct simulations to evaluate the empirical type I error rates under some particular parameter configurations. Specifically, we consider $g$=2, 3, 4, 5, sample size $m_1=\cdots=m_g=n_1=\cdots=n_g$= 20, 40, 60, 80, 100, $\pi_0 = 0.5, 0.6, 0.7, 0.8$, and $\rho_0 = 0.4, 0.5, 0.6$. We also consider some cases with unequal sample sizes: $(m_1,\dots,m_g)=(n_1,\dots,n_g)=(20,40),(20,30,40),(20,25,30,35),(20,25,30,35,40)$ for $g=2,3,4,5$, respectively. So here we consider 288 scenarios. In each configuration, simulations are replicated 50,000 times based on the null hypothesis and the empirical type I error rate of the proposed three methods and Donner's adjusted Chi-square approach are reported. For each of the replication, the observed data $\tilde{D}=(m_{01}, m_{11},m_{21},\ldots, m_{0g}, m_{1g}, m_{2g}, n_{01}, n_{11},\ldots,  n_{0g}, n_{1g})$ are generated from $ (m_{0i},m_{1i},m_{2i})\sim Multinomial(m_i,(R_0{\pi_0}^2-2\pi_0+1, 2\pi_0(1-R_0{\pi_0}), R_0{\pi_0}^2))$ and $ n_{1i}\sim Binomial(n_i,\pi_0), i=1,\dots,g $, where $R_0=\frac{(1-\pi_0)\rho_0}{ \pi_0}+1$. Here, we assume equal dependence (i.e. equal R) between two eyes of the same person across groups, while in practice, R model may not be correct if the equal dependence assumption does not hold in the data set. Based on the generated data, we estimate the MLEs of $\pi_i, i=1,\dots,g$ and $R$ under null hypothesis and alternative hypothesis. Then the test statistics of the three proposed testing procedures can be calculated based on 
the MLEs derived in Section 2. We reject the null hypothesis $H_0: \pi_1=\pi_2=\cdots =\pi_g$ if the estimated p-value is less than or equal to 0.05. The empirical type I error rates are calculated as {\it the number of rejections}$/50 000$. The results are presented in Table ~\ref{tabSimuAlpha}.
\subsubsection{Empirical type I error under the whole parameter space}\hfill\\
In addition to the above specific parameter settings, given the number of groups $g$ and sample size, we also generate parameters $\pi_0$ and $\rho_0$ randomly from $Uniform(0,1)$. Specifically, we consider $g$=2, 3, 4, 5, sample size $m_1=\cdots=m_g=n_1=\cdots=n_g$= 20, 40, 60, 80, 100. In each design, 1000 pairs of $\pi_0$ and $\rho_0$ are generated independently from $Uniform(0,1)$, for each of the 1000 pairs, empirical type I error rates are computed as described in Section 3.1.1. We present the 1000 empirical type I error rates under each design in a boxplot, results from the 20 scenarios are presented in the 20 bosplots in Figure 1-5.
\subsubsection{Powers}\hfill\\
Next, we evaluate the performance of powers for the proposed methods. We consider the alternative hypotheses with
$H_1: \pi=$(0.25, 0.4), (0.25, 0.325, 0.4), (0.25, 0.3, 0.35, 0.4) and (0.25, 0.29, 0.33, 0.37, 0.4) for $g$=2, 3, 4, 5, respectively.
We choose $R$ as 1, 1.5, 2.0 and sample size as $m_1=\cdots=m_g=n_1=\cdots=n_g=$ 20, 40, 60, 80, 100. We also consider some cases with unequal sample sizes: $(m_1,\dots,m_g)=(n_1,\dots,n_g)=(20,40),(20,30,40),(20,25,30,35),(20,25,30,35,40)$ and $(m_1,\dots,m_g)=(n_1,\dots,n_g)=(60,80),(60,70,80), (60,65,70,75),(60,65,70,75,80)$ for $g=2,3,4,5$, respectively. For each of the design, following the same process to compute empirical type I error as described in Section 3.1.1, we compute the power under the alternative hypothesis. The results are presented in Table~\ref{tabSimuPower}.

\subsection{Simulation results}\hfill\\
The simulation results for empirical type I error are displayed in Table ~\ref{tabSimuAlpha} and Figure 1-5. Table ~\ref{tabSimuAlpha} shows the empirical type I error under some specific parameter settings. To simplify the notation, let $(m_1,\dots,m_g)=m$ and $(n_1,\dots,n_g)=n$ for the equal sample size scenarios. Following \citet{Tang_2008}, we say a test is liberal if the ratio of its actual type I error rate to the nominal type I error rate is greater than 1.2 (e.g., type I error rate $>$ 0.06 for $\alpha$ = 0.05, in bold in Table ~\ref{tabSimuAlpha}), conservative if the ratio is less than 0.8 (e.g., type I error rate $<$ 0.04), and robust otherwise. Figure 1-5 show 20 boxplots, each boxplot presents the results of 1000 empirical type I error rates for the 1000 pairs of $\pi_0$ and $\rho_0$ generated from $Uniform(0,1)$ under the given sample size and number of groups ($g$). According to Table ~\ref{tabSimuAlpha} and Figure 1-5, for $m=n=20,40$ and $g=4,5$, LR tests produce some liberal results while Wald tests are extremely liberal; when $g>2$, Wald tests are more liberal than LR tests. As $g$ becomes larger, Wald tests become more liberal under the same sample size. Under the unequal sample size scenarios, almost all empirical type I error rates for Wald tests are greater than 0.06. In addition, as shown in Table ~\ref{tabSimuAlpha}, the empirical type I error rates for Wald tests decrease with the increase of the correlation coefficient ($\rho$) for $g=2$, while for larger number of groups (e.g. $g$=5), Wald tests become more liberal as the increase of correlation coefficient. There is no effect of correlation coefficient on empirical type I error rate for these tests except for Wald tests. Also, there is no effect of $\pi_i$ on empirical type I error rate for all the four tests. The results for Donner's adjusted chi-square test do not change too much under different configurations. According to Figure 1-5, Score tests and Donner's approach are always robust under different configurations with the median near the nominal type I error rate. In general, the proposed three tests get closer to the nominal level when sample size goes larger. 

The simulation results for statistical power under the specific parameter settings are presented in Table~\ref{tabSimuPower}. LR and Wald tests are more powerful than score tests and Donner's procedures under the given configurations. However, LR and Wald tests do not produce satisfactory type I error control because their empirical levels are larger than the nominal level (see Table~\ref{tabSimuAlpha}) under some situations, so we do not recommend LR and Wald tests. When two eyes from the same patient are completely independent (i.e. $R=1$), score test is as powerful as Donner's test; when two eyes from the same patient are not completely independent (i.e. $R\neq1$), score test is more powerful than Donner's test; when $R=2$, score test is remarkably powerful than Donner's test. So generally, score test yields more powerful results compared with Donner's test, thus we highly recommend score test.

\begin{figure}[htpb]
\centering
 \includegraphics[width=18cm,height=9.5cm]{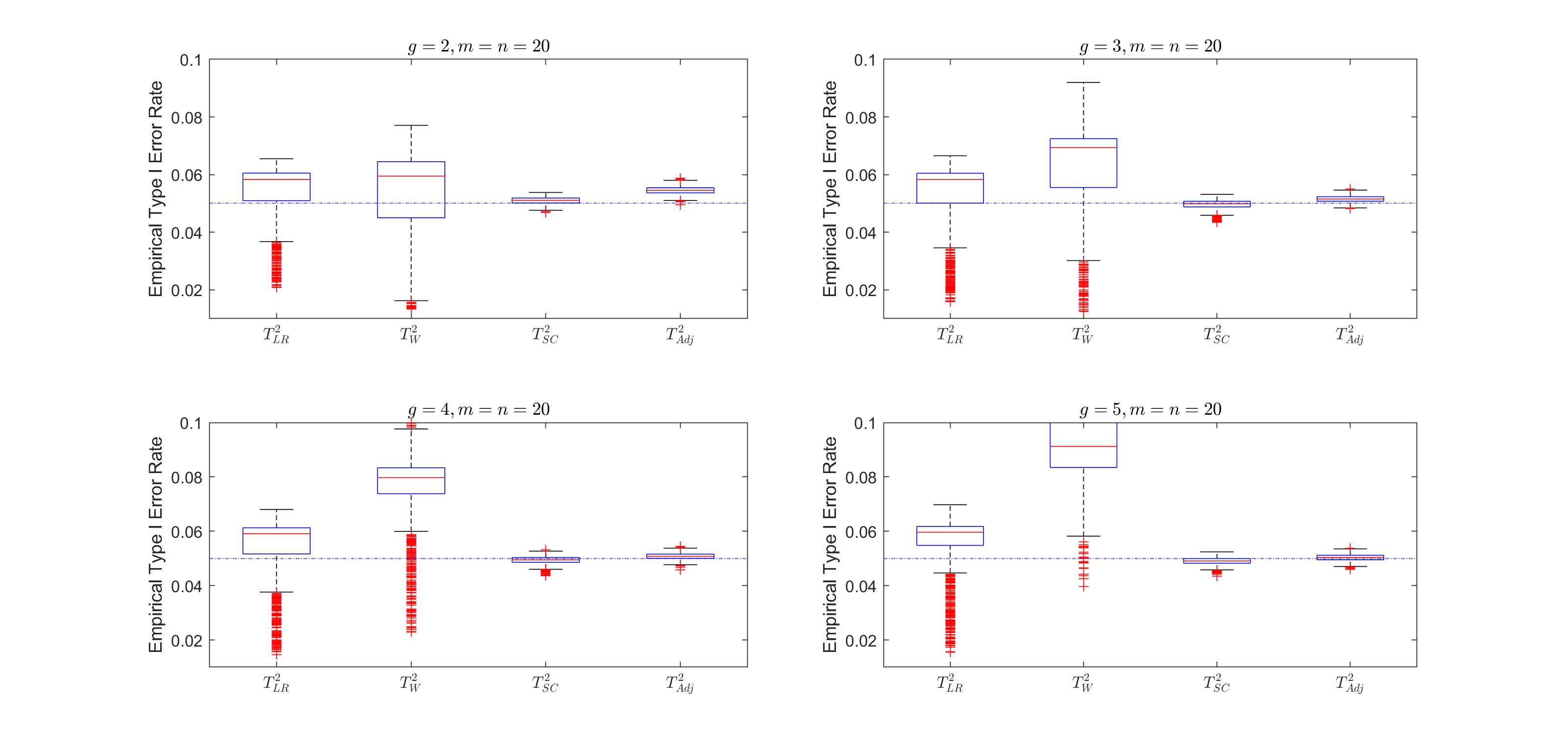}
 \caption{Boxplots for empirical type I error rates ($m=n=20$).}

\end{figure}

\begin{figure}[htpb]
\centering
 \includegraphics[width=18cm,height=9.5cm]{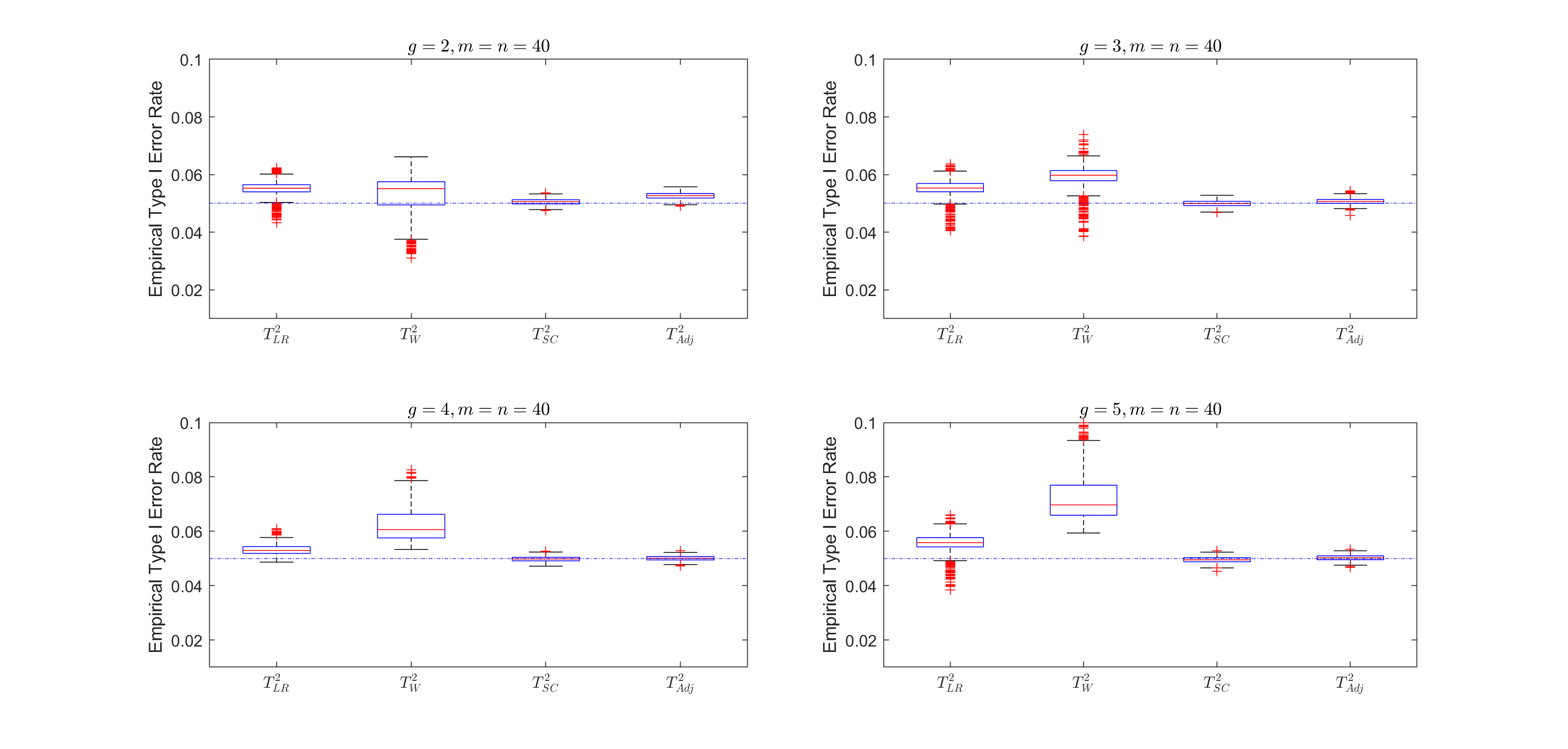}
 \caption{Boxplots for empirical type I error rates ($m=n=40$).}

\end{figure}

\begin{figure}[htpb]
\centering
 \includegraphics[width=18cm,height=9.5cm]{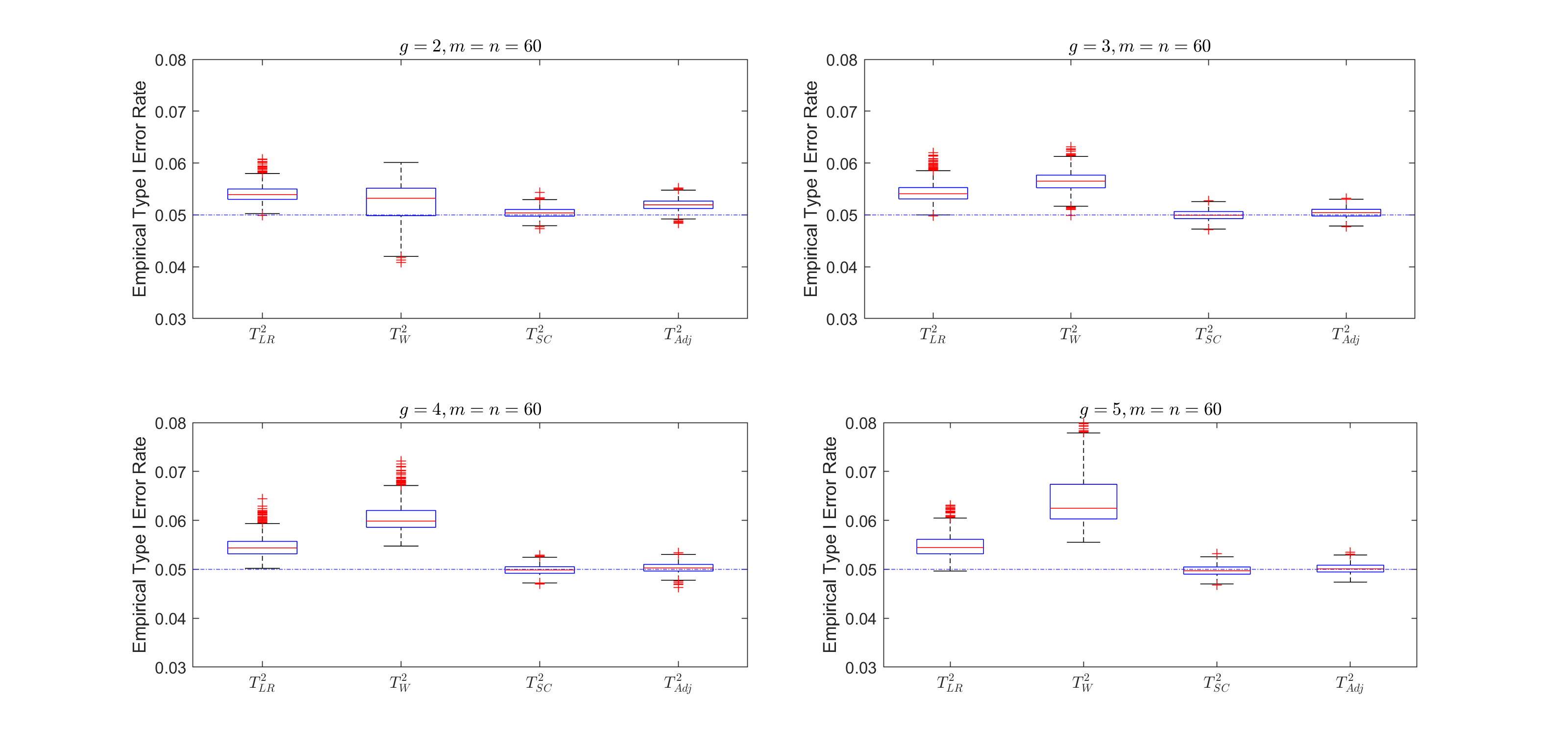}
 \caption{Boxplots for empirical type I error rates ($m=n=60$).}

\end{figure}

\begin{figure}[htpb]
\centering
 \includegraphics[width=18cm,height=9.5cm]{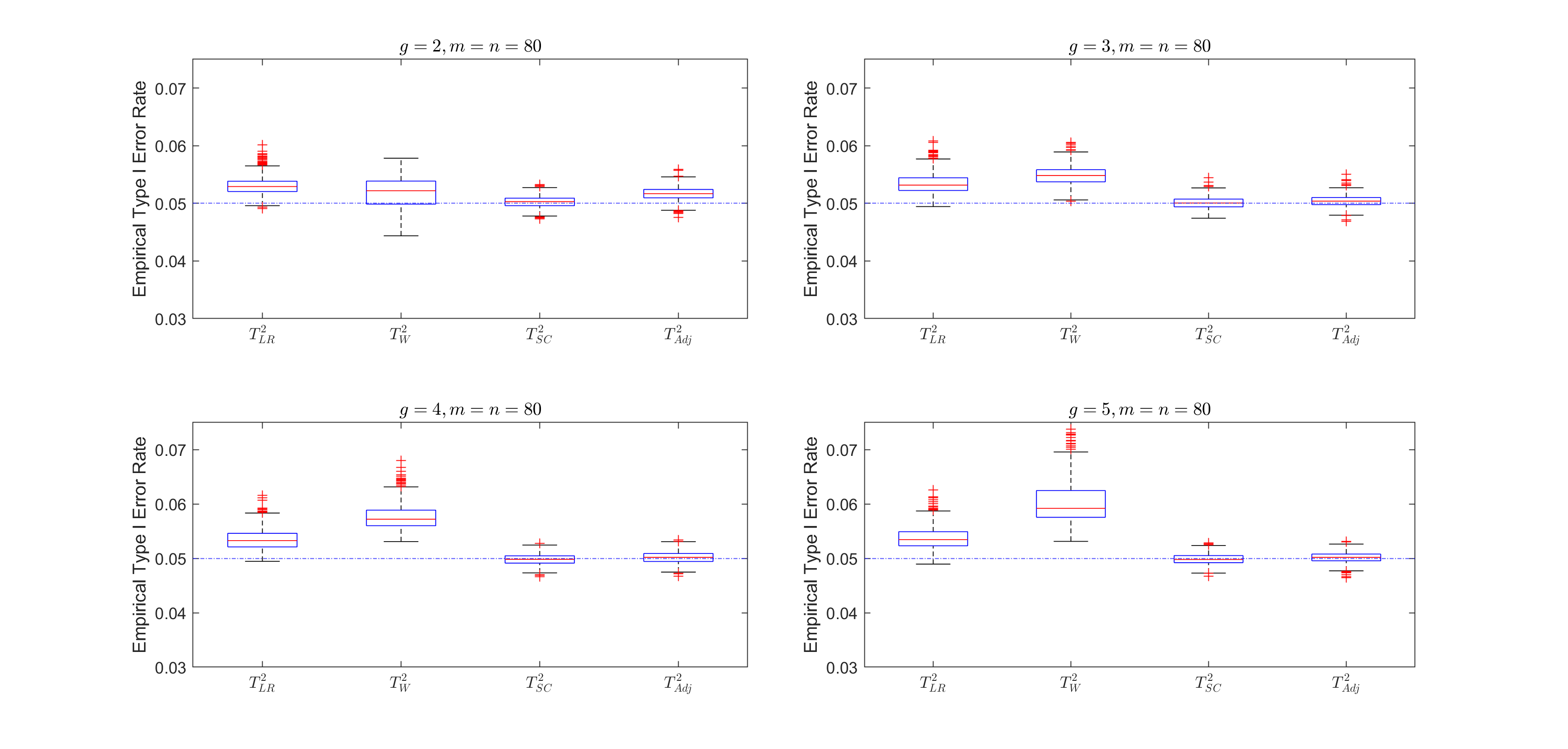}
 \caption{Boxplots for empirical type I error rates ($m=n=80$).}

\end{figure}

\begin{figure}[htpb]
\centering
 \includegraphics[width=18cm,height=9.5cm]{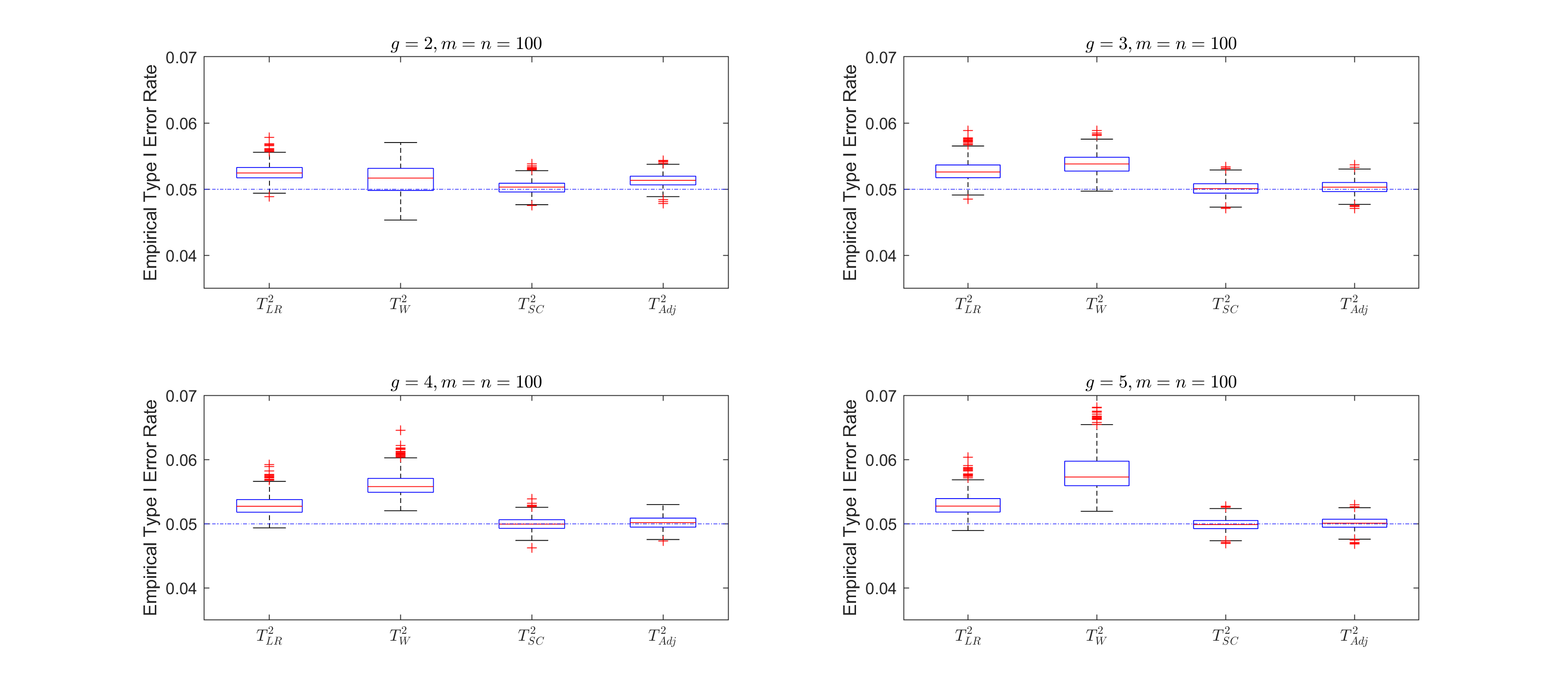}
 \caption{Boxplots for empirical type I error rates ($m=n=100$).}

\end{figure}

\renewcommand{\arraystretch}{0.7}

{\small\tabcolsep=3pt

\begin{longtable}{cccccccccccccccccccccccc}

\caption{\label{tabSimuAlpha}The empirical type I error rates (percent) of various testing procedures under $H_0: \pi_1=\cdots=\pi_g=\pi_0$ at $\alpha=0.05$ based on 50,000 replicates}\\
\hline
$m$&$n$&$\pi_0$&$\rho$&&\multicolumn{4}{c}{$g=2$}&&\multicolumn{4}{c}{$g=3$}&&\multicolumn{4}{c}{$g=4$}&&\multicolumn{4}{c}{$g=5$}\\
\cline{6-9}\cline{11-14}\cline{16-19}\cline{20-24}
&&&&& $T_{LR}^2$ & $T_W^2$ & $T_{SC}^2$ & $T_{Adj}^2$&&$T_{LR}^2$ & $T_W^2$ & $T_{SC}^2$ & $T_{Adj}^2$&&$T_{LR}^2$ & $T_W^2$ & $T_{SC}^2$ & $T_{Adj}^2$&&$T_{LR}^2$ & $T_W^2$ & $T_{SC}^2$ & $T_{Adj}^2$\\
\hline
\endfirsthead
\multicolumn{24}{c}%
{\tablename\ \thetable\  \textit{Continued from previous page}} \\
\hline
$m$&$n$&$\pi_0$&$\rho$&&\multicolumn{4}{c}{$g=2$}&&\multicolumn{4}{c}{$g=3$}&&\multicolumn{4}{c}{$g=4$}&&\multicolumn{4}{c}{$g=5$}\\
\cline{6-9}\cline{11-14}\cline{16-19}\cline{20-24}
&&&&& $T_{LR}^2$ & $T_W^2$ & $T_{SC}^2$ & $T_{Adj}^2$&&$T_{LR}^2$ & $T_W^2$ & $T_{SC}^2$ & $T_{Adj}^2$&&$T_{LR}^2$ & $T_W^2$ & $T_{SC}^2$ & $T_{Adj}^2$&&$T_{LR}^2$ & $T_W^2$ & $T_{SC}^2$ & $T_{Adj}^2$\\
\hline
\endhead
\hline \multicolumn{24}{r}{\textit{Continued on next page}} \\
\endfoot
\hline \multicolumn{24}{l}{\textit{a: unequal sample sizes: $(m_1,\dots,m_g)=(n_1,\dots,n_g)=(20,40),(20,30,40),(20,25,30,35),(20,25,30,35,40)$ for $g=2,3,$}}\\
 \multicolumn{24}{l}{\textit{ $4,5$, respectively.}}\\
\endlastfoot
20 & 20 & 0.5 & 0.4 &  &5.88  & 6.00  & 5.00  & 5.14  &  & 5.87  & \textbf {6.82} & 4.94  & 5.26  &  & \textbf {6.15} & \textbf {7.88} & 5.08  & 5.13  &  & \textbf {6.16} & \textbf {8.72} & 4.96  & 5.07    \\ 
       &  &  & 0.5 &  &5.80  & 5.44  & 5.06  & 5.07  &  & 5.77  & \textbf {6.61} & 4.90  & 5.14  &  & \textbf {6.08} & \textbf {8.11} & 5.04  & 5.05  &  & \textbf {6.31} & \textbf {9.35} & 4.92  & 5.06    \\ 
       &  &  & 0.6 &  &5.38  & 4.56  & 5.08  & 5.26  &  & 5.20  & 5.59  & 4.97  & 5.31  &  & 5.57  & \textbf {7.98} & 5.06  & 5.07  &  & 5.83  & \textbf {10.30} & 4.89  & 5.00    \\ 
       &  & 0.6 & 0.4 &  &5.69  & 5.92  & 4.86  & 5.25  &  & \textbf {6.01} & \textbf {7.01} & 5.19  & 5.27  &  & 5.95  & \textbf {7.82} & 4.95  & 5.07  &  & \textbf {6.21} & \textbf {8.95} & 5.07  & 5.10    \\ 
       &  &  & 0.5 &  &5.50  & 5.25  & 4.98  & 5.33  &  & 5.78  & \textbf {6.71} & 5.12  & 5.22  &  & 5.85  & \textbf {7.94} & 4.87  & 5.04  &  & \textbf {6.16} & \textbf {9.45} & 5.07  & 5.11    \\ 
       &  &  & 0.6 &  &4.91  & 4.16  & 4.98  & 5.22  &  & 4.86  & 5.61  & 5.10  & 5.31  &  & 5.15  & \textbf {8.07} & 4.91  & 4.95  &  & 5.50  & \textbf {10.96} & 4.90  & 5.16    \\ 
       &  & 0.7 & 0.4 &  &5.57  & 5.76  & 5.08  & 5.16  &  & 5.93  & \textbf {7.18} & 5.06  & 5.14  &  & 5.97  & \textbf {8.12} & 4.87  & 4.94  &  & \textbf {6.12} & \textbf {9.05} & 5.03  & 4.87    \\ 
       &  &  & 0.5 &  &5.23  & 4.82  & 5.02  & 5.18  &  & 5.15  & \textbf {6.57} & 4.95  & 5.06  &  & 5.52  & \textbf {8.57} & 4.84  & 5.09  &  & 5.82  & \textbf {10.32} & 4.95  & 4.88    \\ 
       &  &  & 0.6 &  &4.41  & 3.64  & 5.03  & 5.23  &  & 4.08  & 5.24  & 4.94  & 5.16  &  & 4.51  & \textbf {8.59} & 4.89  & 4.95  &  & 5.02  & \textbf {12.52} & 4.93  & 4.89    \\ 
       &  & 0.8 & 0.4 &  &4.83  & 5.02  & 4.86  & 5.25  &  & 5.25  & \textbf {7.34} & 5.01  & 5.05  &  & 5.55  & \textbf {9.60} & 4.68  & 4.96  &  & \textbf {6.05} & \textbf {11.70} & 4.87  & 4.91    \\ 
       &  &  & 0.5 &  &4.08  & 3.82  & 4.89  & 5.19  &  & 4.14  & \textbf {6.42} & 4.83  & 4.96  &  & 4.61  & \textbf {10.19} & 4.69  & 4.98  &  & 5.06  & \textbf {14.03} & 4.73  & 4.92    \\ 
       &  &  & 0.6 &  &3.06  & 2.50  & 4.91  & 5.15  &  & 2.83  & 4.47  & 4.65  & 4.99  &  & 3.22  & \textbf {9.16} & 4.60  & 5.01  &  & 3.61  & \textbf {14.96} & 4.67  & 4.82    \\ 
      40 & 40 & 0.5 & 0.4 &  &5.57  & 5.67  & 5.11  & 5.28  &  & 5.56  & \textbf {6.02} & 5.04  & 5.02  &  & 5.69  & \textbf {6.39} & 5.16  & 5.10  &  & 5.51  & \textbf {6.55} & 4.99  & 4.99    \\ 
       &  &  & 0.5 &  &5.43  & 5.20  & 4.99  & 5.18  &  & 5.62  & 5.80  & 5.12  & 5.00  &  & 5.68  & \textbf {6.42} & 5.00  & 4.99  &  & 5.58  & \textbf {6.80} & 4.92  & 4.97    \\ 
       &  &  & 0.6 &  &5.50  & 4.94  & 4.95  & 5.28  &  & 5.65  & 5.74  & 4.95  & 5.02  &  & 5.75  & \textbf {6.51} & 5.01  & 4.97  &  & 5.83  & \textbf {7.13} & 5.01  & 4.96    \\ 
       &  & 0.6 & 0.4 &  &5.54  & 5.54  & 5.14  & 5.27  &  & 5.58  & 5.94  & 5.09  & 5.10  &  & 5.66  & \textbf {6.46} & 5.09  & 5.16  &  & 5.67  & \textbf {6.77} & 5.09  & 5.11    \\ 
       &  &  & 0.5 &  &5.61  & 5.31  & 5.11  & 5.23  &  & 5.52  & 5.85  & 4.99  & 5.11  &  & 5.86  & \textbf {6.50} & 5.21  & 5.14  &  & 5.80  & \textbf {7.05} & 5.14  & 5.04    \\ 
       &  &  & 0.6 &  &5.55  & 4.98  & 5.06  & 5.14  &  & 5.68  & 5.82  & 5.12  & 5.16  &  & 5.74  & \textbf {6.50} & 5.01  & 5.10  &  & \textbf {6.04} & \textbf {7.38} & 5.24  & 5.00    \\ 
       &  & 0.7 & 0.4 &  &5.34  & 5.30  & 4.93  & 5.05  &  & 5.66  & 5.98  & 5.20  & 5.18  &  & 5.66  & \textbf {6.46} & 5.20  & 4.92  &  & 5.54  & \textbf {6.92} & 4.95  & 5.03    \\ 
       &  &  & 0.5 &  &5.32  & 5.06  & 4.90  & 5.03  &  & 5.63  & 5.97  & 5.04  & 5.09  &  & 5.85  & \textbf {6.58} & 5.21  & 5.00  &  & 5.72  & \textbf {7.09} & 4.99  & 5.03    \\ 
       &  &  & 0.6 &  &5.42  & 4.84  & 4.98  & 5.04  &  & 5.67  & 5.93  & 5.04  & 5.07  &  & 5.83  & \textbf {6.89} & 5.05  & 4.92  &  & 5.84  & \textbf {7.53} & 5.02  & 5.00    \\ 
       &  & 0.8 & 0.4 &  &5.44  & 5.33  & 4.99  & 5.10  &  & 5.71  & \textbf {6.09} & 5.13  & 5.05  &  & 5.67  & \textbf {6.64} & 5.03  & 4.94  &  & 5.64  & \textbf {7.33} & 4.97  & 4.95    \\ 
       &  &  & 0.5 &  &5.45  & 5.17  & 5.00  & 5.15  &  & 5.51  & \textbf {6.07} & 4.87  & 5.02  &  & 5.70  & \textbf {7.05} & 5.03  & 5.00  &  & 5.83  & \textbf {7.93} & 4.98  & 5.07    \\ 
       &  &  & 0.6 &  &5.26  & 4.62  & 5.01  & 5.06  &  & 5.11  & \textbf {6.16} & 4.87  & 5.02  &  & 5.40  & \textbf {7.52} & 5.00  & 4.91  &  & 5.61  & \textbf {8.79} & 4.99  & 5.06    \\ 
      60 & 60 & 0.5 & 0.4 &  &5.34  & 5.35  & 5.09  & 5.29  &  & 5.43  & 5.74  & 5.14  & 5.05  &  & 5.49  & 5.99  & 5.14  & 5.02  &  & 5.69  & \textbf {6.40} & 5.24  & 5.21    \\ 
       &  &  & 0.5 &  &5.34  & 5.16  & 5.07  & 5.26  &  & 5.48  & 5.68  & 5.11  & 5.07  &  & 5.53  & 5.99  & 5.10  & 5.02  &  & 5.56  & \textbf {6.31} & 5.20  & 5.16    \\ 
       &  &  & 0.6 &  &5.30  & 4.90  & 4.94  & 5.27  &  & 5.63  & 5.57  & 5.18  & 5.06  &  & 5.59  & \textbf {6.12} & 5.07  & 5.00  &  & 5.69  & \textbf {6.50} & 5.19  & 5.09    \\ 
       &  & 0.6 & 0.4 &  &5.32  & 5.30  & 5.05  & 5.18  &  & 5.31  & 5.50  & 5.02  & 5.09  &  & 5.45  & \textbf {6.02} & 5.13  & 5.15  &  & 5.35  & 5.96  & 4.98  & 5.10    \\ 
       &  &  & 0.5 &  &5.37  & 5.19  & 5.10  & 5.30  &  & 5.36  & 5.45  & 4.98  & 5.00  &  & 5.44  & 5.93  & 5.07  & 5.09  &  & 5.35  & \textbf {6.01} & 4.94  & 5.14    \\ 
       &  &  & 0.6 &  &5.33  & 4.93  & 5.01  & 5.19  &  & 5.37  & 5.39  & 4.95  & 5.08  &  & 5.37  & 5.82  & 4.94  & 5.08  &  & 5.37  & \textbf {6.17} & 4.85  & 5.15    \\ 
       &  & 0.7 & 0.4 &  &5.37  & 5.30  & 5.10  & 5.28  &  & 5.47  & 5.73  & 5.14  & 5.09  &  & 5.29  & 5.83  & 4.98  & 5.01  &  & 5.39  & \textbf {6.11} & 4.99  & 5.16    \\ 
       &  &  & 0.5 &  &5.43  & 5.15  & 5.14  & 5.31  &  & 5.53  & 5.66  & 5.18  & 5.20  &  & 5.44  & 5.91  & 5.10  & 4.99  &  & 5.45  & \textbf {6.24} & 5.02  & 5.04    \\ 
       &  &  & 0.6 &  &5.42  & 5.03  & 5.07  & 5.28  &  & 5.56  & 5.68  & 5.11  & 5.11  &  & 5.49  & \textbf {6.09} & 5.00  & 4.99  &  & 5.53  & \textbf {6.44} & 4.93  & 5.08    \\ 
       &  & 0.8 & 0.4 &  &5.23  & 5.10  & 4.93  & 5.21  &  & 5.41  & 5.66  & 5.03  & 5.03  &  & 5.34  & 5.95  & 4.95  & 4.95  &  & 5.40  & \textbf {6.30} & 4.96  & 5.14    \\ 
       &  &  & 0.5 &  &5.34  & 5.08  & 5.03  & 5.16  &  & 5.55  & 5.74  & 5.08  & 4.90  &  & 5.27  & \textbf {6.08} & 4.79  & 4.94  &  & 5.44  & \textbf {6.64} & 4.90  & 5.11    \\ 
       &  &  & 0.6 &  &5.36  & 5.00  & 4.97  & 5.26  &  & 5.48  & 5.66  & 4.96  & 4.99  &  & 5.26  & \textbf {6.32} & 4.62  & 5.00  &  & 5.53  & \textbf {7.17} & 4.91  & 5.13    \\ 
      80 & 80 & 0.5 & 0.4 &  &5.36  & 5.36  & 5.11  & 5.25  &  & 5.21  & 5.42  & 4.93  & 5.16  &  & 5.37  & 5.75  & 5.08  & 5.11  &  & 5.28  & 5.77  & 5.02  & 4.92    \\ 
       &  &  & 0.5 &  &5.32  & 5.22  & 5.06  & 5.26  &  & 5.35  & 5.44  & 5.05  & 5.19  &  & 5.38  & 5.72  & 5.05  & 5.27  &  & 5.32  & 5.79  & 4.97  & 4.97    \\ 
       &  &  & 0.6 &  &5.28  & 4.95  & 5.02  & 5.25  &  & 5.36  & 5.34  & 5.02  & 5.04  &  & 5.48  & 5.77  & 5.05  & 5.19  &  & 5.34  & 5.84  & 4.97  & 4.89    \\ 
       &  & 0.6 & 0.4 &  &5.34  & 5.33  & 5.15  & 5.39  &  & 5.32  & 5.55  & 5.09  & 5.15  &  & 5.34  & 5.66  & 5.04  & 5.08  &  & 5.24  & 5.87  & 4.98  & 5.03    \\ 
       &  &  & 0.5 &  &5.35  & 5.19  & 5.15  & 5.34  &  & 5.38  & 5.45  & 5.16  & 5.16  &  & 5.22  & 5.51  & 5.00  & 4.98  &  & 5.39  & 5.99  & 5.01  & 5.02    \\ 
       &  &  & 0.6 &  &5.27  & 5.00  & 5.00  & 5.30  &  & 5.38  & 5.37  & 5.04  & 5.13  &  & 5.27  & 5.60  & 5.02  & 5.00  &  & 5.49  & \textbf {6.07} & 5.15  & 4.97    \\ 
       &  & 0.7 & 0.4 &  &5.34  & 5.26  & 5.17  & 5.24  &  & 5.08  & 5.30  & 4.85  & 5.08  &  & 5.22  & 5.61  & 4.97  & 5.00  &  & 5.31  & 5.83  & 4.98  & 5.15    \\ 
       &  &  & 0.5 &  &5.34  & 5.17  & 5.11  & 5.23  &  & 5.15  & 5.35  & 4.85  & 4.96  &  & 5.27  & 5.67  & 4.95  & 5.00  &  & 5.36  & 5.96  & 4.99  & 5.11    \\ 
       &  &  & 0.6 &  &5.32  & 5.00  & 5.08  & 5.20  &  & 5.22  & 5.40  & 4.94  & 5.03  &  & 5.49  & 5.93  & 5.06  & 5.02  &  & 5.36  & \textbf {6.16} & 4.96  & 5.10    \\ 
       &  & 0.8 & 0.4 &  &5.31  & 5.24  & 5.07  & 5.22  &  & 5.36  & 5.51  & 5.06  & 5.13  &  & 5.41  & 5.77  & 5.14  & 5.03  &  & 5.35  & \textbf {6.00} & 5.04  & 5.09    \\ 
       &  &  & 0.5 &  &5.39  & 5.19  & 5.13  & 5.29  &  & 5.40  & 5.59  & 5.08  & 5.06  &  & 5.52  & \textbf {6.05} & 5.18  & 5.02  &  & 5.54  & \textbf {6.42} & 5.16  & 5.20    \\ 
       &  &  & 0.6 &  &5.43  & 5.06  & 5.08  & 5.24  &  & 5.31  & 5.50  & 4.85  & 5.14  &  & 5.64  & \textbf {6.20} & 5.19  & 5.06  &  & 5.53  & \textbf {6.64} & 5.02  & 5.18    \\ 
      100 & 100 & 0.5 & 0.4 &  &5.20  & 5.18  & 5.04  & 5.15  &  & 5.34  & 5.45  & 5.14  & 5.12  &  & 5.20  & 5.48  & 5.02  & 5.01  &  & 5.26  & 5.70  & 5.00  & 5.06    \\ 
       &  &  & 0.5 &  &5.20  & 5.09  & 5.06  & 5.19  &  & 5.47  & 5.53  & 5.27  & 5.17  &  & 5.35  & 5.52  & 5.12  & 4.97  &  & 5.26  & 5.65  & 4.99  & 5.03    \\ 
       &  &  & 0.6 &  &5.16  & 4.93  & 4.98  & 5.07  &  & 5.42  & 5.37  & 5.19  & 5.10  &  & 5.29  & 5.60  & 5.02  & 4.97  &  & 5.32  & 5.69  & 5.06  & 4.94    \\ 
       &  & 0.6 & 0.4 &  &5.30  & 5.32  & 5.14  & 5.15  &  & 5.19  & 5.33  & 5.02  & 4.95  &  & 5.07  & 5.38  & 4.91  & 5.06  &  & 5.24  & 5.60  & 5.05  & 5.05    \\ 
       &  &  & 0.5 &  &5.36  & 5.23  & 5.20  & 5.20  &  & 5.22  & 5.25  & 5.01  & 5.06  &  & 5.14  & 5.46  & 4.91  & 4.93  &  & 5.18  & 5.57  & 4.95  & 5.01    \\ 
       &  &  & 0.6 &  &5.27  & 5.04  & 5.07  & 5.14  &  & 5.28  & 5.27  & 5.04  & 5.03  &  & 5.19  & 5.42  & 4.94  & 4.89  &  & 5.33  & 5.76  & 4.96  & 4.86    \\ 
       &  & 0.7 & 0.4 &  &5.22  & 5.20  & 5.09  & 5.12  &  & 5.28  & 5.44  & 5.10  & 5.05  &  & 5.20  & 5.39  & 5.02  & 4.99  &  & 5.18  & 5.53  & 4.95  & 5.05    \\ 
       &  &  & 0.5 &  &5.26  & 5.14  & 5.09  & 5.02  &  & 5.19  & 5.30  & 4.95  & 5.13  &  & 5.19  & 5.46  & 4.96  & 4.92  &  & 5.21  & 5.68  & 4.95  & 4.97    \\ 
       &  &  & 0.6 &  &5.28  & 5.05  & 5.08  & 5.03  &  & 5.37  & 5.37  & 5.13  & 5.03  &  & 5.29  & 5.64  & 5.03  & 4.88  &  & 5.30  & 5.88  & 5.00  & 4.98    \\ 
       &  & 0.8 & 0.4 &  &5.30  & 5.24  & 5.16  & 5.06  &  & 5.36  & 5.40  & 5.15  & 5.08  &  & 5.20  & 5.55  & 5.03  & 5.12  &  & 5.12  & 5.64  & 4.94  & 4.92    \\ 
       &  &  & 0.5 &  &5.25  & 5.09  & 5.05  & 5.10  &  & 5.40  & 5.47  & 5.15  & 5.04  &  & 5.32  & 5.60  & 5.05  & 5.11  &  & 5.22  & 5.86  & 4.93  & 5.03    \\ 
       &  &  & 0.6 &  &5.24  & 4.99  & 4.99  & 5.18  &  & 5.39  & 5.43  & 5.05  & 5.07  &  & 5.40  & 5.91  & 5.07  & 5.16  &  & 5.37  & \textbf {6.18} & 5.05  & 5.03    \\ 
  
       a & a & 0.5 & 0.4 &  &5.73  & \textbf {6.29} & 5.05  & 5.23  &  & 5.90  & \textbf {6.85} & 5.09  & 5.14  &  & 5.91  & \textbf {7.29} & 5.04  & 5.02  &  & 5.88  & \textbf {7.84} & 4.97  & 5.08    \\ 
       &  &  & 0.5 &  &5.64  & \textbf {6.17} & 5.08  & 5.31  &  & 5.90  & \textbf {6.83} & 5.19  & 5.05  &  & 5.97  & \textbf {7.57} & 4.99  & 5.01  &  & 5.95  & \textbf {8.20} & 4.94  & 5.03    \\ 
       &  &  & 0.6 &  &5.33  & 5.88  & 5.07  & 5.25  &  & 5.79  & \textbf {6.89} & 5.10  & 5.10  &  & 5.93  & \textbf {7.93} & 5.11  & 5.06  &  & \textbf {6.06} & \textbf {8.97} & 4.93  & 5.01    \\ 
       &  & 0.6 & 0.4 &  &5.69  & \textbf {6.27} & 5.15  & 5.35  &  & 5.79  & \textbf {6.69} & 5.11  & 5.02  &  & 5.74  & \textbf {7.24} & 4.99  & 5.07  &  & 5.81  & \textbf {7.70} & 5.07  & 5.16    \\ 
       &  &  & 0.5 &  &5.67  & \textbf {6.26} & 5.21  & 5.29  &  & 5.75  & \textbf {6.80} & 4.98  & 5.00  &  & 5.90  & \textbf {7.38} & 4.99  & 5.08  &  & 5.91  & \textbf {8.19} & 5.00  & 5.18    \\ 
       &  &  & 0.6 &  &5.24  & \textbf {6.05} & 5.11  & 5.34  &  & 5.52  & \textbf {6.91} & 4.98  & 5.01  &  & 5.61  & \textbf {8.00} & 4.97  & 5.08  &  & \textbf {6.03} & \textbf {8.97} & 4.95  & 5.02    \\ 
       &  & 0.7 & 0.4 &  &5.62  & \textbf {6.22} & 5.10  & 5.17  &  & 5.59  & \textbf {6.72} & 4.90  & 4.97  &  & 5.92  & \textbf {7.57} & 5.15  & 5.12  &  & \textbf {6.02} & \textbf {8.07} & 5.03  & 5.12    \\ 
       &  &  & 0.5 &  &5.38  & \textbf {6.17} & 5.06  & 5.14  &  & 5.62  & \textbf {6.90} & 5.00  & 4.99  &  & 5.85  & \textbf {7.88} & 5.06  & 5.05  &  & 5.94  & \textbf {8.59} & 5.01  & 5.03    \\ 
       &  &  & 0.6 &  &4.79  & \textbf {6.00} & 5.06  & 5.11  &  & 5.12  & \textbf {7.38} & 4.81  & 5.02  &  & 5.39  & \textbf {9.06} & 5.11  & 5.13  &  & 5.76  & \textbf {10.05} & 4.96  & 5.06    \\ 
       &  & 0.8 & 0.4 &  &5.44  & \textbf {6.51} & 5.04  & 5.26  &  & 5.56  & \textbf {7.19} & 4.81  & 4.89  &  & 5.91  & \textbf {8.58} & 4.91  & 5.12  &  & 5.90  & \textbf {9.11} & 4.93  & 5.02    \\ 
       &  &  & 0.5 &  &4.79  & \textbf {6.35} & 4.92  & 5.19  &  & 5.12  & \textbf {7.79} & 4.61  & 4.92  &  & 5.47  & \textbf {9.61} & 4.90  & 5.05  &  & 5.88  & \textbf {10.60} & 4.99  & 5.06    \\ 
       &  &  & 0.6 &  &3.96  & \textbf {6.08} & 4.87  & 5.24  &  & 4.26  & \textbf {8.75} & 4.69  & 4.91  &  & 4.40  & \textbf {11.61} & 4.69  & 4.96  &  & 5.10  & \textbf {14.61} & 4.82  & 4.98    \\ 

\end{longtable}

}

{\small\tabcolsep=3pt
\begin{table}\caption{\label{tabSimuPower}The powers (percent) of various testing procedures at $\alpha=0.05$ based on 50,000 replicates}
\begin{tabular}{ccccccccccccccccccccccc}\hline
$m$&$n$&$R$&&\multicolumn{4}{c}{$g=2$}&&\multicolumn{4}{c}{$g=3$}&&\multicolumn{4}{c}{$g=4$}&&\multicolumn{4}{c}{$g=5$}\\
\cline{5-8}\cline{10-13}\cline{15-18}\cline{20-23}
&&&& $T_{LR}^2$&$T_{W}^2$ & $T_{SC}^2$ & $T_{Adj}^2$&&$T_{LR}^2$&$T_{W}^2$ & $T_{SC}^2$& $T_{Adj}^2$&&$T_{LR}^2$&$T_{W}^2$ & $T_{SC}^2$& $T_{Adj}^2$&&$T_{LR}^2$&$T_{W}^2$ & $T_{SC}^2$& $T_{Adj}^2$\\\hline
 20 & 20 & 1.0 &  & 42.5  &44.4  & 40.6  & 42.6  &  & 33.9  & 36.8  & 31.9  & 33.0 &  & 32.2  & 36.1  & 30.2  & 31.2 &  & 32.9  & 37.6  & 30.7  & 31.6  \\ 
       &  & 1.5 &  & 40.2  &41.6  & 36.9  & 37.5  &  & 32.1  & 34.6  & 28.7  & 28.4 &  & 30.5  & 34.2  & 27.2  & 26.7 &  & 31.1  & 36.0  & 27.5  & 27.0  \\ 
       &  & 2.0 &  & 45.6  &42.8  & 41.9  & 33.3  &  & 37.3  & 36.7  & 33.7  & 24.9 &  & 35.6  & 37.3  & 31.9  & 23.5 &  & 37.3  & 40.8  & 32.6  & 23.5  \\ 
      40 & 40 & 1.0 &  & 70.5  &71.4  & 69.7  & 70.5  &  & 60.1  & 61.7  & 59.0  & 59.7 &  & 58.6  & 60.6  & 57.4  & 58.1 &  & 60.8  & 63.3  & 59.6  & 60.1  \\ 
       &  & 1.5 &  & 67.0  &67.8  & 65.2  & 63.7  &  & 56.6  & 58.2  & 54.6  & 52.6 &  & 54.9  & 57.2  & 52.8  & 50.8 &  & 56.7  & 59.4  & 54.6  & 52.5  \\ 
       &  & 2.0 &  & 74.7  &73.6  & 72.3  & 58.0  &  & 65.3  & 65.1  & 62.3  & 46.6 &  & 64.4  & 65.0  & 61.1  & 45.0 &  & 66.5  & 67.8  & 63.1  & 46.0  \\ 
      60 & 60 & 1.0 &  & 86.0  &86.4  & 85.6  & 86.0  &  & 79.0  & 79.7  & 78.4  & 78.8 &  & 77.7  & 78.8  & 77.1  & 77.5 &  & 80.6  & 81.7  & 80.0  & 80.3  \\ 
       &  & 1.5 &  & 83.0  &83.5  & 82.1  & 80.3  &  & 75.5  & 76.4  & 74.2  & 71.9 &  & 74.0  & 75.2  & 72.8  & 70.1 &  & 76.8  & 78.2  & 75.4  & 72.6  \\ 
       &  & 2.0 &  & 89.1  &88.6  & 88.0  & 74.6  &  & 83.4  & 83.3  & 81.6  & 65.6 &  & 82.7  & 83.1  & 80.9  & 63.4 &  & 85.3  & 85.7  & 83.4  & 65.9  \\ 
      80 & 80 & 1.0 &  & 94.1  &94.3  & 94.0  & 94.2  &  & 89.4  & 89.8  & 89.2  & 89.3 &  & 89.3  & 89.7  & 89.0  & 89.1 &  & 91.1  & 91.6  & 90.9  & 90.9  \\ 
       &  & 1.5 &  & 92.2  &92.4  & 91.8  & 90.3  &  & 86.6  & 87.2  & 85.9  & 83.8 &  & 86.4  & 87.0  & 85.7  & 83.4 &  & 88.5  & 89.2  & 87.7  & 85.7  \\ 
       &  & 2.0 &  & 96.0  &95.8  & 95.5  & 86.3  &  & 92.6  & 92.6  & 91.8  & 78.0 &  & 92.7  & 92.8  & 91.7  & 77.6 &  & 94.2  & 94.4  & 93.4  & 80.0  \\ 
      100 & 100 & 1.0 &  & 97.6  &97.7  & 97.6  & 97.6  &  & 95.1  & 95.3  & 95.0  & 95.1 &  & 95.1  & 95.3  & 95.0  & 95.1 &  & 96.5  & 96.7  & 96.4  & 96.5  \\ 
       &  & 1.5 &  & 96.5  &96.6  & 96.4  & 95.4  &  & 93.3  & 93.6  & 93.0  & 91.4 &  & 93.3  & 93.6  & 93.0  & 91.4 &  & 94.9  & 95.2  & 94.6  & 93.1  \\ 
       &  & 2.0 &  & 98.5  &98.5  & 98.4  & 92.7  &  & 96.9  & 96.9  & 96.5  & 87.3 &  & 97.0  & 97.1  & 96.7  & 86.7 &  & 97.9  & 97.9  & 97.6  & 89.1  \\ 
  a & a & 1.0 &  & 53.2  &56.2  & 50.7  & 52.1  &  & 44.3  & 48.4  & 41.9  & 42.9 &  & 40.5  & 45.0  & 38.2  & 39.0 &  & 44.0  & 49.2  & 41.2  & 42.0  \\ 
       &  & 1.5 &  & 50.0  &53.1  & 44.6  & 45.0  &  & 41.7  & 45.6  & 37.0  & 36.4 &  & 38.0  & 42.6  & 33.7  & 33.0 &  & 40.8  & 46.8  & 35.9  & 35.3  \\ 
       &  & 2.0 &  & 57.8  &55.8  & 51.0  & 39.6  &  & 50.3  & 49.9  & 44.3  & 31.5 &  & 46.1  & 47.5  & 40.2  & 28.5 &  & 50.1  & 52.4  & 43.9  & 30.2  \\ 
b & b & 1.0 &  & 90.5  &91.0  & 90.2  & 90.3  &  & 84.3  & 85.2  & 83.8  & 84.1 &  & 82.6  & 83.6  & 82.1  & 82.3 &  & 85.9  & 87.0  & 85.3  & 85.6  \\ 
       &  & 1.5 &  & 87.9  &88.6  & 86.9  & 85.6  &  & 81.1  & 82.2  & 79.8  & 77.4 &  & 79.2  & 80.5  & 77.9  & 75.1 &  & 82.6  & 84.1  & 81.2  & 78.5  \\ 
       &  & 2.0 &  & 92.8  &92.5  & 91.7  & 80.5  &  & 88.6  & 88.5  & 86.9  & 71.1 &  & 87.4  & 87.5  & 85.7  & 68.5 &  & 90.0  & 90.3  & 88.4  & 72.0  \\

\hline
$H_1:$&$\pi$=&&&  \multicolumn{4}{c}{(0.25, 0.4)} &&  \multicolumn{4}{c}{(0.25,0.325,0.4)}&&  \multicolumn{4}{c}{(0.25,0.3,0.35,0.4)}&&  \multicolumn{4}{c}{(0.25,0.29,0.33,0.37,0.4)}\\
%\multicolumn{17}{l}{Note: $T$ is the test statistic in Rosner (1982).}
\hline  \multicolumn{23}{l}{a: unequal sample sizes: $(m_1,\dots,m_g)=(n_1,\dots,n_g)=(20,40),(20,30,40),(20,25,30,35),(20,25,30,35,40)$ for} \\ \multicolumn{23}{l}{ $g=2,3,4,5$, respectively. b: unequal sample sizes: $(m_1,\dots,m_g)=(n_1,\dots,n_g)=(60,80),(60,70,80),$ }\\ \multicolumn{23}{l}{  $(60,65,70,75),(60,65,70,75,80)$ for $g=2,3,4,5$, respectively.}
\end{tabular}
\end{table}
}

\section{A work example}
We revisit the example mentioned in Section 1, the clinical trial was conducted to compare cefaclor and amoxicillin for the treatment of acute otitis media with effusion (OME) after the tympanocentesis. 214 children aged 2 months - 16 years underwent unilateral or bilateral tympanocentesis and then were assigned to receive one of those two antibiotics randomly\cite{Mandel_1982}. The sample used for this analysis consisted of 173 children out of the sample of 214 children with 93 receiving Cefaclor and 80 receiving Amoxicillin (Table~\ref{tabExmData42}). Here, instead of using data from the presence or absence of OME at 14 days after initiation of treatment, we use data from the presence or absence of OME at 42 days in children related to OME status at entry. Following the notation given in Section 2, Table~\ref{tabExmData42} can also be shown as in Table~\ref{tabExmDataforus} with $m_{01}=9, m_{11}=7, m_{21}=23, m_{02}=7, m_{12}=5, m_{22}=13, n_{01}=20, n_{11}=34, n_{02}=19, n_{12}=36$. We are interested in testing the equality of proportions of ears being cured in the two treatment groups. In the original study, the conclusion was given by only describing the percentage of children without effusion or "improved" in the two treatment groups (68.9\% vs. 67.5\%), without testing the treatment effect. Also, the intra-class correlations were not taken into consideration. For the rest of this section, we apply the proposed methods in Section 2 and compare it with the results in the original study.

The testing procedures developed in this article are based on Rosner's R model (i.e. equal dependence between two eyes across groups). Thus, before applying the proposed methods to this example, we perform goodness of fit tests to examine whether R model is appropriate for the analysis of this data set. \citet{liu2020goodness} developed and explored several goodness-of-fit tests to tackle the intra-class correlation problem arising in bilateral data, according to their findings, likelihood ratio test ($G^2$) and Pearson chi-square test ($\chi^2$) work well for Rosner's model, which give us $G^2=0.3871$ with p-value= 0.5338 and $\chi^2= 0.3867$ with p-value= 0.5341. Both methods indicate that Rosner's model fit the data well, so we can apply the proposed testing procedures to this data set.

Based on the data given above, we obtain MLEs of parameters and p-values and test statistics of the three proposed tests and Donner's test. An overall significant difference between the proportion of ears without OME in the two groups is shown in Table~\ref{tabExmStat}. Table~\ref{tabMLE} shows the constrained MLEs and unconstrained MLEs of the proportion of ears without OME $\pi_i$, the correlation coefficient $\rho_i$, $i=1,2$ in the two treatment groups and the intra-class dependence measurement $R$. The unconstrained MLEs of $\pi_1$ and $\pi_2$ are  0.6528 and 0.6425, respectively, which are slightly lower than the the percentage of children whose ears were without effusion or "improved" in the two treatment groups (68.9\% vs. 67.5\%) in the original study. The p-values for the three proposed methods and Donner's statistics are all greater than 0.05, indicating that we fail to reject the null hypothesis: $H_0: \pi_1=\pi_2$, which corresponds to the result in the original paper that by 42 days after entry the percentage of children without OME was equal in both treatment groups. However, by performing the statistical tests taking into account the intra-class correlation, our methods provide much stronger statistical evidence that the rate of ears being cured was equal in the two treatment groups.       

%We use a double-blind randomized clinical trial in the paper by Mandel et al. \cite{Mandel_1982} to illustrate the newly proposed methods. The clinical trial was conducted to compare cefaclor and amoxicillin for the treatment of acute otitis media with effusion (OME) after the tympanocentesis. 214 children aged 2 months - 16 years underwent unilateral or bilateral tympanocentesis and then were assigned to receive a 14-day course of one of those two antibiotics randomly. The sample used for this analysis consisted of 203 children out of the sample of 214 children with 106 receiving Cefaclor and 97 receiving Amoxicillin (Table~\ref{tabExmData}).

%An overall significant difference between the proportion of ears without OME in the two groups is from 0.2967 to 0.4569 based on the proposed methods and 0.0261 on Donner's statistic (Table~\ref{tabExmStat}).

\begin{table*}
\centering
\caption{\label{tabExmData42}Presence or absence of OME at 42 days in children related to OME status at entry}

\begin{threeparttable}
\begin{tabular}{ccccccccc}  \hline
&&\multicolumn{7}{c}{OME Status at 42 Days}\\\cline{3-9}
Entry OME Status  &&\multicolumn{3}{c}{Cefaclor (N=93)} && \multicolumn{3}{c}{Amoxicillin (N=80)}   \\\cline{3-5} \cline{7-9} 
&&No OME& Unilateral  &Bilateral &&No OME &Unilateral &Bilateral \\&&&OME&OME&&&OME&OME\\
%&&No OME& Unilateral OME  &Bilateral OME&&No OME&Unilateral OME&Bilateral OME
\hline

 Unilateral OME &&  34 &  20& &&36&19&  \\
 
 Bilateral OME&&   23  & 7 & 9 && 13 & 5 &7\\\hline 

   % {ccccccccc}{p{3cm}p{1cm}p{1.5cm}p{1.5cm}p{1.5cm}p{1cm}p{1.5cm}p{1.5cm}p{1.5cm}}
 \end{tabular}
 
      \end{threeparttable}
\end{table*}

\begin{table*}
\caption{\label{tabExmDataforus}Distribution of the number of ears without disease at 42 Days}
\begin{threeparttable}
\begin{tabular}{cccc}  \hline
&&\multicolumn{2}{c}{Treatment Group}\\\cline{3-4}
Number of ears being cured  & &  Cefaclor     &  Amoxicillin  \\\hline
 0 &&  9  &  7\\
 1 &&  7   &  5 \\
 2 &&   23   &  13 \\
 total &&   39   &  25 \\\hline 
 0 &&  20   & 19 \\
 1 &&   34   & 36 \\
 total &&   54   & 55 \\\hline
 
 \end{tabular}
 
      \end{threeparttable}
\end{table*}

\begin{table}
\caption{\label{tabExmStat}Statistics and p-values for comparing the proportion of ears without OME for different groups}
\begin{tabular}{ccccc}\hline
Method & $T_{LR}^2$ & $T_W^2$ & $T_{SC}^2$ & $T_{Adj}^2$ \\\hline

p-value & 0.8426  &0.8432 &0.8424 &0.7688\\
statistic &0.0394 &0.0391&0.0395 &0.0864 \\\hline
\end{tabular}\end{table}

\begin{table*}
\caption{\label{tabMLE}Constrained and Unconstrained MLEs}
\begin{threeparttable}
\begin{tabular}{ccccccc}  \hline
&&\multicolumn{2}{c}{Unconstrained MLEs}&&\multicolumn{2}{c}{Constrained MLEs}\\\cline{3-4} \cline{6-7}
MLE  & &  Cefaclor     &  Amoxicillin& &  Cefaclor     &  Amoxicillin  \\\hline
 $\hat{\pi}_i$ &&  $\hat{\pi}_1=0.6528$  & $\hat{\pi}_2=0.6425$&&  \multicolumn{2}{c}{$\hat{\pi}_{H_0}=0.6482$}\\
 $\hat{R}$ && \multicolumn{2}{c}{$\hat{R}=1.3172$}&&  \multicolumn{2}{c}{$\hat{R}_{H_0}=1.3182$} \\
 $\hat{\rho}$ && $\hat{\rho}_1=0.5964$   &  $\hat{\rho}_2= 0.5699 $ &&  \multicolumn{2}{c}{$\hat{\rho}_{H_0}=0.5862$} \\\hline 

 \end{tabular}
 
      \end{threeparttable}
\end{table*}

\section{Conclusions}
In this article, we derived three procedures for testing the equality of proportions for combined correlated bilateral and unilateral data under Rosner's model. We investigated the performance of proposed methods by exploring empirical type I error rates and powers in simulation studies and applied to an example. We also provided comparisons to Donner's adjusted Chi-square approach. The proposed procedures are based on likelihood estimate algorithm derived by utilizing the root of 4th order polynomial equations and fisher scoring method, which is very efficient since only $R$ is updated by the fisher scoring iterations. %In addition, the asymptotic testing procedures are given so we can calculate the exact value of the test statistics and get the empirical type I error rates in simulation studies.

Score test is recommended because it has both satisfactory empirical type I error rates and powers under different number of groups, sample sizes and parameter configurations. While LR test and Wald test have inflated type I error rates under small sample size. Wald test is more liberal than LR test, especially for larger number of group. The performance of the proposed three tests becomes more similar as sample size goes larger.

The methods proposed in this article may be applied to other areas of medical research, where correlated bilateral data and unilateral data are observed on paired organs of the same individual. One assumption under Rosner's model is equal measurements of dependence between two eyes of the same person (i.e., equal R across all groups). Thus, one should perform goodness of fit tests to examine whether the equal R assumption holds before using Ronsner's model. Due to asymptotic properties of the proposed tests, approaches in this article do not perform well for small sample size scenarios, in these situations, exact test is needed to overcome the inflated type I error rate problem. We consider exact test for $g>2$ as a future work. Our approaches, which provide explicit forms of test statistics and MLEs, can improve the computational efficiency, and therefore can be useful for the development of exact test.

\clearpage

\medskip

\bibliographystyle{unsrtnat}
\bibliography{main}

\clearpage

\section*{Appendix}

\subsection{Information matrix}

Differentiating ${\partial l\over\partial \pi_i}, i=1,\ldots, g$ and ${\partial l\over\partial R}$ with respect to $\pi_i, i=1,\ldots, g$ and $R$ respectively yields

\begin{eqnarray*}
  {\partial^2 l\over\partial \pi_i^2} &=& \frac{m_{0i}\, \left( - 2\, R^2\, \pi_i^2 + 4\, R\, \pi_i + 2\, R - 4\right)}{{\left(R\, \pi_i^2 - 2\, \pi_i + 1\right)}^2} - \frac{2\, m_{2i}}{\pi_i^2} - \frac{\left(2\, R^2\, \pi_i^2 - 2\, R\, \pi_i + 1\right)\, m_{1i}}{\pi_i^2\, {\left(R\, \pi_i - 1\right)}^2} 
 - \frac{n_{1i}}{\pi_i^2} + \frac{n_{0i}}{(1-\pi_i)^2}, \\
  {\partial^2 l\over\partial \pi_i\partial R} &=& - \frac{m_{1i}}{{\left(R\, \pi_i - 1\right)}^2}
  - \frac{2\, \left(\pi_i - 1\right)\, \pi_i m_{0i}}{{\left(R\, {\pi_i}^2 - 2\, \pi_i + 1\right)}^2}, \\
  &&\ \ \ \ \ \  i=1, \ldots, g\\
  {\partial^2 l\over\partial \pi_i\partial \pi_j} &=& 0, i\ne j, \\
  {\partial^2 l\over\partial R^2} &=&   - \frac{S_2}{R^2} - \sum_{i=1}^g \frac{\pi_i^2\, m_{1i}}{{\left(R\, \pi_i - 1\right)}^2} - \sum_{i=1}^g \frac{\pi_i^4\, m_{0i}}{{\left(R\, \pi_i^2 - 2\, \pi_i + 1\right)}^2}.
\end{eqnarray*}
Then we have
\begin{eqnarray*}
  I_{ii}&=&E\left(-{\partial^2 l\over\partial \pi_i^2}\right) = \frac{2\, m_i\, \left(2\, R^2\, {\pi_i}^2 - R\, {\pi_i}^2 - 2\, R\, \pi_i + 1\right)}{\pi_i\, {\left(R\, {\pi_i}^2 - 2\, \pi_i + 1\right)}{\left(1 - R\, \pi_i\right)}}
 + \frac{n_{i}}{\pi_i(1-\pi_i)}, \\
  I_{i,g+1}&=&E\left(-{\partial^2 l\over\partial \pi_i\partial R}\right) = -\frac{2\, \left(1 - R\right)\, {\pi_i}^2\, m_i}{{\left(R\, {\pi_i}^2 - 2\, \pi_i + 1\right)}{\left(1 - R\, \pi_i\right)}}, \\
  &&\ \ \ \ \ \  i=1, \ldots, g\\
  I_{ij}&=&E\left(-{\partial^2 l\over\partial \pi_i\partial \pi_j}\right) = 0, i\ne j, \\
  I_{g+1,g+1}&=&E\left(-{\partial^2 l\over\partial R^2}\right) =  \sum_{i=1}^g\frac{{\pi_i}^2\, m_i (R\pi_i - 2\, \pi_i + 1)}{R{\left(R\, {\pi_i}^2 - 2\, \pi_i + 1\right)}{\left(1 - R\, \pi_i\right)}}.
\end{eqnarray*}
The $(g+1)\times(g+1)$ information matrix is denoted as $I(\pi_1, \ldots, \pi_g; R)=(I_{ij})$.

It is straightforward but tedious to show that the inverse of information matrix can be expressed as
$$I^{-1}(\pi; R)= 
\begin{bmatrix}
   c_{11} & c_{12} & \cdots & c_{1g} & d_1 \\
   c_{21} & c_{22} & \cdots & c_{2d} & d_2 \\
  \cdots & \cdots & \cdots & \cdots & \cdots \\
  c_{g1} & c_{g2} & \cdots & c_{gg} & d_g \\
  d_1 & d_2 & \cdots & d_g & f \\
\end{bmatrix}
$$
where
\begin{eqnarray*}
f &=& \left(I_{g+1,g+1}-\sum_{k=1}^g \frac{I_{k,g+1}^2}{I_{kk}} \right)^{-1} \\
c_{ii} &=&  \frac{1}{I_{ii}} + \frac{I_{i, g+1}^2 f}{I_{ii}^2}, i=1, \cdots, g\\
c_{ij} &=&  \frac{I_{i, g+1}I_{j, g+1} f}{I_{ii}I_{jj}}, i\neq j\\
d_{i} &=&   -\frac{I_{i, g+1} f}{I_{ii}}, i=1, \cdots, g.
\end{eqnarray*}

\end{document}